\documentclass{jfm}
\usepackage{amsmath}
\usepackage{graphicx}
\graphicspath{{images/}}
\usepackage{natbib}
\usepackage{physics}
\usepackage{listings}
\usepackage[toc, page]{appendix}

\title{Signature of coalescence during scalar mixing in a rankine vortex}

\author{Sabyasachi Sen\aff{1}, 
    Prajwal\aff{1}, 
    Joris Heyman\aff{2},
    Tanguy Le Borgne\aff{2}, 
 \and Aditya Bandopadhyay\aff{1} \corresp{\email{aditya@mech.iitkgp.ac.in}}}

\affiliation{\aff{1}Department of Mechanical Engineering, Indian Institute of Technology Kharagpur, Kharagpur,West Bengal - 721302, India
\aff{2}G\'eosciences Rennes, UMR 6118, Universit\'e de Rennes 1, CNRS, 35000 Rennes, France}

\begin{document}

\maketitle

\begin{abstract}    We analyze the dynamics of solute mixing in a vortex flow. 
The transport of a passive tracer is considered in a Rankine vortex. The action of a shear flow, in general, is to achieve stretching of fluid elements. A vortex flow exhibits stretching and folding of fluid elements in a way which brings adjacent fluid elements closer every turn. A strong stretching along the direction of rotation is accompanied by a concomitant thinning in the radial direction leading to a strong diffusive flux which may cause material from neighbouring regions of the mixing interface to aggregate. Through a Lagrangian concentration evolution technique, the diffusive strip method, we obtain the concentration field and pinpoint the signature of coalescence of two neighbouring concentration regions by analyzing the concentration distribution profiles. We link coalescence with reactivity for mixing-limited reactive flows. The analysis is useful to understand scalar dispersion in vortical flow structures. 
\end{abstract}

\section{Introduction}

An accurate description of solute mixing is crucial in order to understand a host of geophysical processes \citep{pierrehumbert1991large, fernando1991turbulent}. Scalar mixing is central to geological storage of carbon dioxide \citep{caldeira2000accelerating}, contaminant transport in groundwater \citep{zheng2002applied}, growth of biofilms \citep{tel2005chemical}, plastic and pharmaceutical processing \citep{mohr1957mixing, nienow1997mixing} ozone layer dynamics and other chemical processes \citep{cerbelli2002prediction, shankar2009mixing, levenspiel1964patterns}. 

Many turbulent atmospheric flows such as those in the ozonosphere consist of large scale stirring phenomenon \citep{wonhas2002diffusivity} leading to the formation of vortices. These vortices of varying sizes interact with each other and strongly influence mixing of dispersed solutes. Vortical structures also play a major role in oceanic transport phenomenon \citep{lindemann2017dynamics}. A number of studies have focused on the diffusion kinetics and space-filling properties of a vortex flow \citep{flohr1997accelerated, bajer2001accelerated, marble1988mixing}. Stirring protocols which deform the substrate, enhance concentration gradients forming fronts for enhanced chemical reaction \citep{de2005procedure, bandopadhyay2018shear, bandopadhyay2017enhanced}. However, a purely advective description, which does not account for molecular diffusion, of scalar transport does not predict mixing and chemical reaction accurately. Capturing diffusion is very important in a mathematical model that seeks to accurately describe mixing since it is the diffusive flux which is ultimately responsible for the homogenization of a solute in the flow field. While the mechanisms causing elongation of material lines are well understood, predicting mixing rates still remains a challenge particularly when there is a reconnection (or aggregation) between several parts of the mixing interface, leading, at large mixing time, to a so-called coalescence regime \citep{villermaux2019mixing, le2015lamellar}. 

Understanding the dynamics of this coalescence regime becomes important in flows of geophysical relevance since coalescence destroys concentration gradients and reduces effective reaction rates. On the other hand, sharp gradients in concentration form fronts that act as hot-spots for chemical reaction \citep{hidalgo2015dissolution, bates2016engine}. Hence, the key challenge when it comes to quantifying mixing lies in providing an accurate picture of the probability density function (PDF) of the solute concentration and identifying the signature of coalescence during mixing. 

In this work, we study this coalescence dynamics numerically through scalar transport in a Rankine vortex, which is typical of large-scale turbulent flows \citep{pierrehumbert1991large}. To solve the transport problem in these flows, we use a Lagrangian method (the diffusive strip method\citep{meunier2010diffusive}). First, we establish the validity of the diffusive strip method as a much faster alternative to an Eulerian method, which requires a comparatively larger computational power to fully resolve the concentration distribution for high P\'eclet number flows. We use this Lagrangian method to reconstruct the scalar concentration fields and to compute the evolution of the distribution of concentrations levels and scalar energy spectrum in time. The signature of coalescence is clearly observed and we assess the influence of coalescence on the reaction kinetics of mixing-limited reactive flows.

\section{Numerical Methods}

Eulerian methods are useful tools to model the advection-diffusion process but require great computational power to completely resolve the concentration field for high P\'eclet number flows. Due to this shortcoming, efforts have been made to develop alternative modelling techniques based on a different paradigm. In order to study mixing and reaction in fluids, Ranz, in his seminal paper, presented a simplified modelling technique based on a Lagrangian frame of reference fixed on the tracer being advected by the flow and a warped time scale which is a function of the stretching of the strip of scalar. This converted the advection-diffusion problem to a 1-D diffusion equation in a local co-ordinate system \citep{ranz1979applications}. A key assumption behind this simplification is that concentration gradients normal to the strip of tracer are much larger than those along the strip and hence, diffusion is essentially restricted to the direction of maximum concentration gradient.

The Diffusive Strip Method \citep{meunier2010diffusive} is a numerical technique for studying scalar mixing in 2-D which builds on the simplified mathematical framework introduced by Ranz. Passive material strips are advected in the velocity field obtained from flow simulations and their positions are computed kinematically. Diffusion can then be accounted for, independent of advection, by inserting diffusive material segments along the strip. The reduction of the problem to 1-D in suitable co-ordinates requires us to compute only the stretching of the strip, thereby rendering the method very fast.

The advection problem lies in solving the following equation numerically: 

\begin{equation}\label{eq:advection}
    \frac{\partial \mathbf{x_{i}}}{\partial t} = 
    \mathbf{v}(\mathbf{x_{i}},t)
\end{equation}
\newline
Here, the flow field $\mathbf{v}(\mathbf{x_{i}},t)$ is known beforehand and is used to calculate the evolution of the tracers in time. The incompressible nature of the flow field imposes the following kinematic constraint at all times (See Figure \ref{fig:initStripLoc}):     

\begin{equation}
    \frac{d(s.\Delta x)}{dt} = 0
\end{equation}

A Taylor expansion in first order around the point $x_{\textbf{i}}$ along with a transformation that shifts the origin to $x_{\textbf{i}}$ allows us to represent the velocity components along the material line and normal to the material line in terms of the time rate of change of the striation thickness as follows:

\begin{equation}
    U = -\frac{\sigma}{s} \frac{ds}{dt} \quad\mathrm{and}\quad V =  \frac{n}{s} \frac{ds}{dt}
\end{equation}{}

The evolution of the concentration field of the scalar is represented by the advection-diffusion equation:

\begin{equation}\label{eq:advDiffusion}
    \frac{\partial c}{\partial t} + \mathbf{v}  \boldsymbol{\cdot} \nabla{c}= D \laplacian{c}
\end{equation}

The conversion of the advection-diffusion equation to a one-dimensional diffusion equation in a local coordinate system is based on the premise that the lamella length scale in the local normal direction gets compressed such that the transverse scalar gradient is much larger than that along the strip. This implies that concentration gradients perpendicular to the strip are essentially driving diffusion, allowing us to modify the advection-diffusion equation to \citep{meunier2010diffusive}:

\begin{equation}\label{eq:simplified}
    \frac{\partial c}{\partial t} + \frac{n}{s}\frac{ds}{dt}\frac{\partial c}{\partial n} = D\pdv[2]{c}{n}
\end{equation}

Ranz suggests a change of variables where the transverse co-ordinate is normalized by the striation thickness $\tilde{n} = n/s$ and time is warped to $\tau (t) =  \frac{1}{s_{0}^{2}Pe} \int_{0}^{t} \rho (t)^{2} dt$  where $\rho = s_{0}/s$ and $Pe =L^{2}/{t_{adv}D}$.This converts Eq. (\ref{eq:simplified}) to a 1-D diffusion equation \citep{ranz1979applications}:

\begin{equation}\label{eq:1D-Diffusion}
    \frac{\partial c}{\partial \tau} = \frac{\partial^2 c}{\partial \tilde{n}^{2}}
\end{equation}

We assume that the strip of solute contains a conservative scalar whose normalized concentration initially has a Gaussian variation in the local normal coordinate defined as: 

\begin{equation}\label{eq:gaussianInit}
    c(\tilde{n}) = c_{0}e^{-\tilde{n}^{2}}
\end{equation}

Eq. (\ref{eq:1D-Diffusion}) with an initial condition given by (\ref{eq:gaussianInit}) can be solved on an infinite domain to yield:

\begin{equation}\label{eq:concPerpendicularToStrip}
    c(\tilde{n},t) = \frac{c_{0}}{\sqrt{1 + 4\tau_{i}(t)}} exp\left(\frac{-\tilde{n}^{2}}{1 + 4\tau_{i}(t)}\right)
\end{equation}

Eq. (\ref{eq:concPerpendicularToStrip}) indicates that all we need to compute during the particle tracking routine are the position \textbf{$x_{i}$}, the warped time $\tau_{i}$ and the lamella thickness $s_{i}$ of the tracers in order to reconstruct the spatial distribution of the scalar c(\textbf{x},t). This is carried out numerically by adding Gaussian ellipses which are centered at the mid-point of the segment connecting consecutive points on the material strip. 

\begin{equation}\label{eq:reconstructedConcField}
    c(\mathbf{x}) = \sum_{i} \frac{c_{0}/1.7726}{\sqrt{1 + 4\tau_{i}(t)}} exp\left(-\frac{[(\mathbf{x-x_{i}})\cdot \mathbf{\sigma_{i}}] ^ {2}}{\Delta l^{2}} -\frac{[(\mathbf{x-x_{i}})\cdot \mathbf{n_{i}}]^{2}}{s_{i}^{2}(1 + 4\tau_{i}(t))} \right)
\end{equation}

where $\mathbf{\sigma_{i}}$ and $\mathbf{n_{i}}$ are the unit vectors along the strip and normal to the strip respectively [See Figure \ref{fig:initStripLoc}].

In order to test the validity of the diffusive strip method as a viable alternative to Eulerian methods, we consider the prototypical case of a linear shear flow to show the excellent agreement  between the results obtained from simulations carried out in the two different modelling frameworks. The flow field $\mathbf{\overrightarrow{V}} = u\hat{i} + v\hat{j}$ is given by:

\begin{equation}\label{eq:linearShear}
    (u,v) = (y,0)
\end{equation}

\begin{figure}
    \centering
    \includegraphics{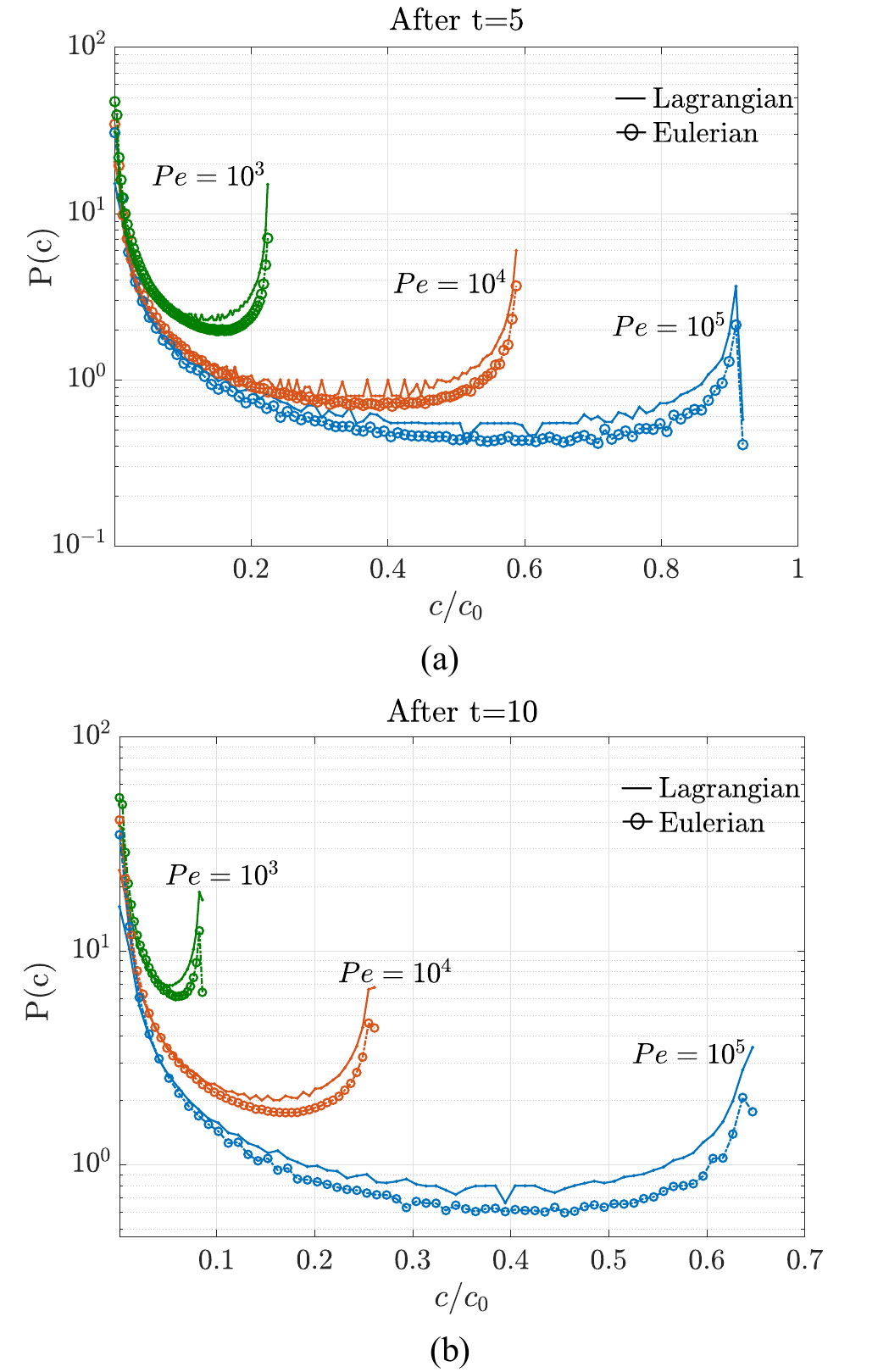}
    \caption{An excellent agreement between the PDFs obtained from the Lagrangian and Eulerian simulations at (a)t=5 and (b)t=10 for a linear shear flow. The solid line corresponds to the Lagrangian simulation whereas the dash-dot line with markers corresponds to the Eulerian simulation.}
    \label{fig:linearShearLagrangianEulerian}
\end{figure}

A strip of scalar of length 1 unit and characteristic thickness 0.1 units is initialized along the y-axis at x=0 and is sheared by this flow field. We reconstruct the concentration field and derive the PDF of concentration using both the Eulerian and Lagrangian schemes. A very good match between the results from the two modelling methods is observed in Figure \ref{fig:linearShearLagrangianEulerian} for a range of P\'eclet numbers. A careful comparison between the PDFs obtained from the Lagrangian and Eulerian simulations reveals that the Eulerian PDFs show a slightly greater decay in concentration than that shown in the results obtained from the diffusive strip method. This additional decay is absent in the Lagrangian simulations precisely because here, the advection and diffusion processes are decoupled. This decoupling causes the velocity field and its gradient to act only on points representing the center-line of the strip and thus the effect of the space-continuous velocity field on the mass comprising the entire material lamella is unaccounted for. This is why we observe uniform stretching normal to the strip in the images for the concentration field reconstructed from the Lagrangian simulations (See Figure \ref{fig:reconstructedConcFieldLagrangian}). This difference is due to the existence of a velocity gradient normal to the strip i.e. $\frac{\partial u}{\partial n} \neq 0$, which causes differential stretching normal to the strip for the case of the Eulerian simulations. 

However, for most of the PDF, the two numerical schemes are in excellent agreement with one another since the differential stretching experienced locally normal to the strip, due to the non-zero contribution from $\frac{\partial u}{\partial n} \Delta n$, is negligible compared to the absolute velocity. This validation exercise establishes the much faster Lagrangian technique based on the diffusive strip method as a very good alternative to Eulerian methods. We proceed in our attempt at identifying signatures of coalescence during mixing with the diffusive strip method as the numerical technique which will yield the actual PDF of concentration of the scalar.

\section{Results}

\subsection{Rankine vortex flow}

The Rankine vortex is a flow model with an inner circular zone comprising a forced vortex and an outer region characterized by a free vortex. The radial symmetry in the flow makes it appropriate to define the velocity field in the cylindrical polar coordinate system (\textrm{r,$\theta$,z}) where the \textrm{z} axis is the axis of symmetry and \textrm{r,$\theta$} coordinates are in the plane of flow. The analytical expression for the Rankine vortex is as follows: 

\begin{eqnarray}
 \left(v_r, v_\theta, v_z \right)= \left\{
    \begin{array}{ll}
          (0,c_{1}r,0), & r \leq r_{c} \\
          (0,\frac{c_{2}}{r},0), & r > r_{c} \\
    \end{array} 
    \right. 
\end{eqnarray}

The velocity field attains a maximum at the characteristic distance of the vortex, \textrm{$r_{c}$}, where the flow field changes its nature from linear to hyperbolic. An important feature of this flow field, with consequences for turbulence \citep{batchelor1953theory}, is a discontinuity in its vorticity, which is a constant $ = 2c_{1}$ in the inner circular zone and a null vector in the outer region. Despite its simplicity, the Rankine vortex has frequently been used to model generic phenomenon in the atmosphere. Using Doppler velocity and reflectivity measurements, it has been shown that the averaged azimuthal velocity distribution in tornadoes \citep{bluestein2003mobile}, mesocyclones \citep{brown2005improved, bertato2003interesting} and some dust devils \citep{bluestein2004doppler} can be modelled very accurately by a Rankine vortex. Since this flow field is so typical of large-scale turbulent atmospheric phenomena, we use it to illustrate coalescence during scalar mixing and also explore some of the consequences of coalescence on the reaction kinetics for mixing-limited reactive flows. 

\begin{figure}
    \centering
    \includegraphics{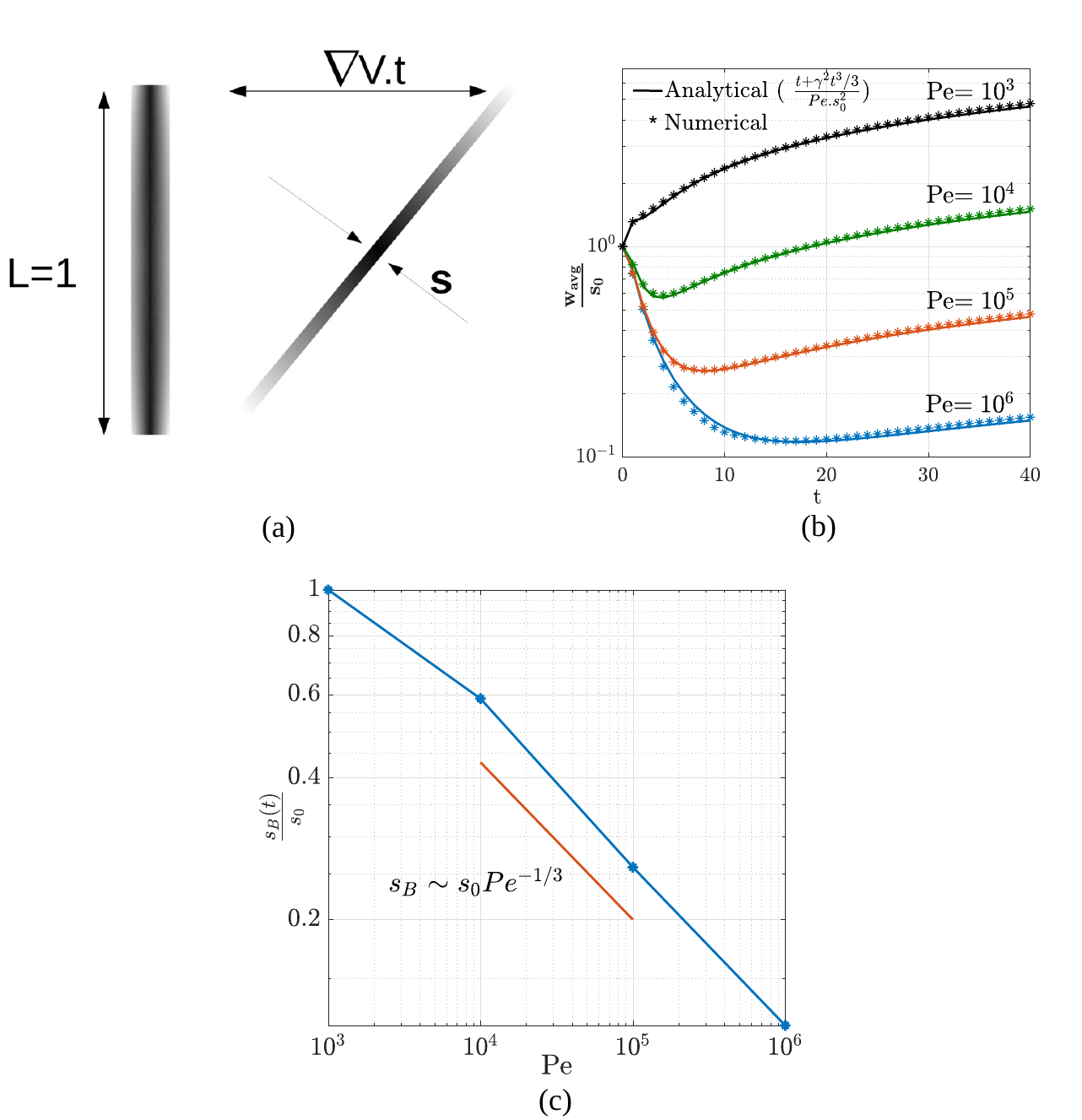}
    \caption{(a)Shearing of a strip of scalar due to the presence of a velocity gradient in the direction perpendicular to the longest dimension of the strip (b) Evolution of strip width in time and (c) Power law scaling between Batchelor scale and P\'eclet number}
    \label{fig:widthVsTime}
\end{figure}

In our numerical study, $c_{1}=2$, $c_{2}=0.5$, $r$ denotes the distance of the material lamella from the center of the vortex and the critical radius where the flow field has a discontinuous slope is $r_{c}=0.5$. All values have been reported in non-dimensional form by assuming the domain length scale to be $O(1)$. A strip of scalar, initially placed along the x-axis between $0.45 < x < 2.0 $, is advected in the flow field of a Rankine vortex for 100 units of time. We assume that in this flow field, the deformation of material lines is governed by a single stretching rate, similar to the case of a linear shear flow (\ref{eq:linearShear}). We denote this stretching rate by $\nabla{v}$ and assume that it is equal to the velocity gradient in the radial direction. Shearing due to this radial velocity gradient increases the length of the strip linearly in time and causes a compression transverse to the strip due to the fact that the velocity field is solenoidal. Figure \ref{fig:widthVsTime}(a) shows this phenomenon yielding the following expression for the striation thickness $s$:

\begin{equation}
    s = \frac{s_{0}}{\sqrt{1+\nabla{v}^{2}t^{2}}}\implies  s \approx \frac{s_{0}}{\nabla{v}t}
    \label{eq:striationThickness}
\end{equation}{}

once shear effects are significant. The thinning of the strip simultaneously increases the concentration gradient transverse to the strip, thereby, inducing a diffusive flux which increases the strip width as $\sqrt{t+t^{3}}$ \citep{meunier2003vortices}. In the initial stages of the vortex wrapping around its center, the competing processes of striation thinning and diffusion broadening result in a net reduction in strip width since shear is the dominant mechanism governing the evolution of the material. The thinning continues until the rate of compression of the strip is equilibrated by the rate of increase in thickness due to diffusion $\frac{s_{0}}{\nabla{v}t} \sim \sqrt{Dt}$. The time at which this event occurs is known as the mixing time \textrm{$t_{mix}$} $\sim \nabla{v}^{-1} Pe^{1/3}$ and the associated length scale is the well-known Batchelor scale \citep{batchelor1959small} $s_{B}$ $\sim s_{0}Pe^{-1/3}$; here, $Pe=(s_{0}^{2}\nabla{v}/D)$. Beyond the mixing time, the rate of compression is not enough to balance the broadening due to diffusion and the strip width increases as $\sqrt{t/Pe}$.

By solving the advection problem (\ref{eq:advection}) and keeping track of the elongation and warped times, we compute the width using $w_{i}=s_{i}\sqrt{1+4\tau_{i}}$ \citep{meunier2010diffusive}. We plot the evolution of this width averaged over the strip, in time, for a range of P\'eclet numbers in Figure \ref{fig:widthVsTime}(b). In the calculation of the analytically averaged width, the value of $\nabla{v}$ is taken to be $0.85$, which is obtained by averaging $\nabla{v} = \abs{\frac{\partial v}{\partial r}}$ over the radial limits of the tracer i.e. $0.45<r<2.0$. The plot shows an excellent agreement between analytical predictions and data from the numerical routine, thus, justifying our assumption that a single stretching rate governs the stirring protocol in the vortex flow. The power law scaling corresponding to the dependence of the Batchelor scale on P\'eclet number also matches well with numerical results (Figure \ref{fig:widthVsTime}(c)).

\subsection{Reconstruction of spatial distribution of scalar}

The task of reconstructing the spatial distribution of the scalar on a 2-D grid is computationally very intensive. An alternative to a complete spatial reconstruction is to derive mixing characteristics from the evolution of 

\begin{equation}\label{eq:cmax1}
    c_{max} = \frac{c_{0}}{\sqrt{1 + 4\tau_{i}(t)}}  
\end{equation} 
and that of 
\begin{equation}
    \rho = \Delta x_{i}/\Delta x_{i}^{0}
\end{equation}
However, this technique will be inaccurate in our case since it doesn't account for the significant overlap between different parts of the strip as it undergoes stretching and folding. Hence, a complete reconstruction becomes essential in order to capture coalescence. 

\begin{figure}
    \centering
    \includegraphics[scale=0.8]{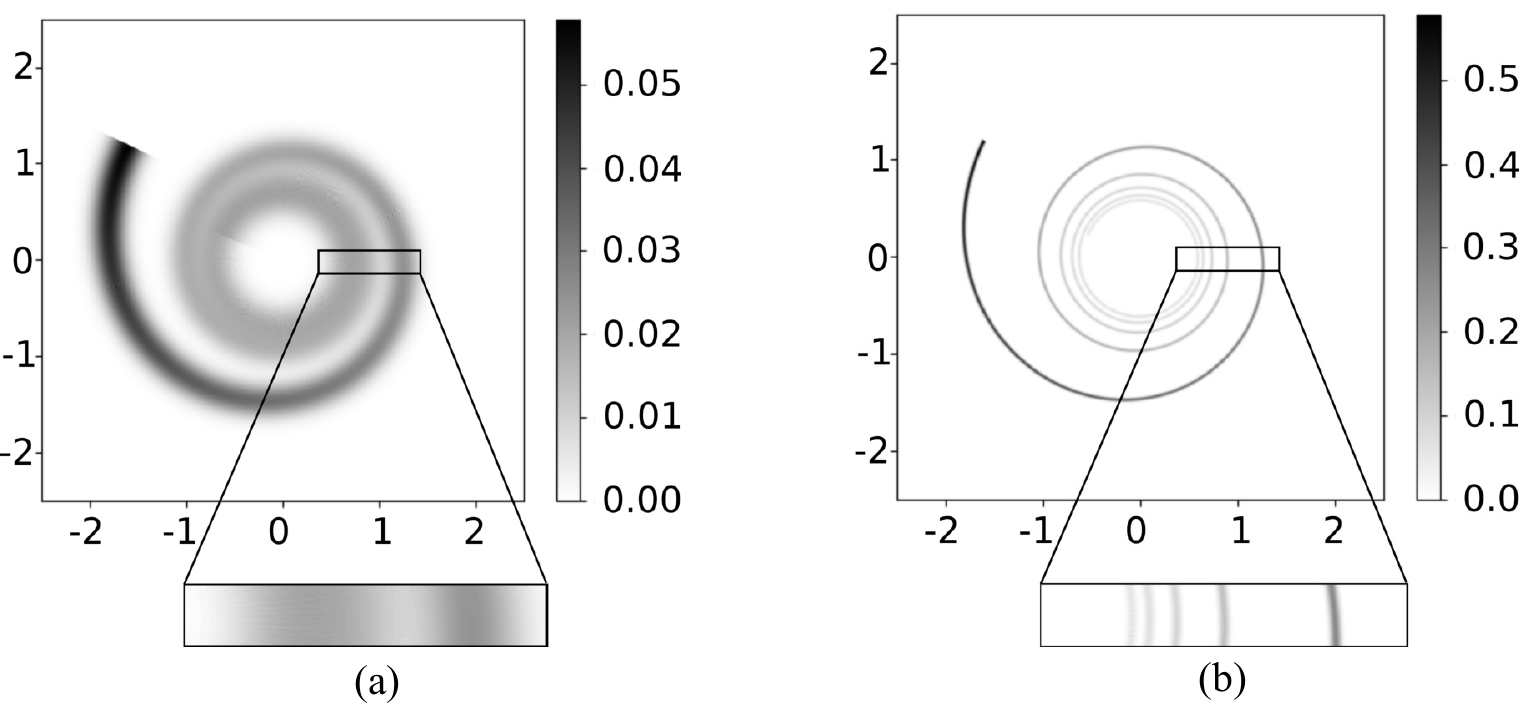}
    \caption{Reconstruction of spatial distribution of scalar at t=20 for Pe = a)$10^{3}$ and d) $10^{5}$}
    \label{fig:reconstructedConcFieldLagrangian}
\end{figure}

We reconstruct the spatial distribution of the scalar on a 2-D grid for two representative P\'eclet numbers using the diffusive strip method as the much faster alternative to fully resolved Eulerian simulations. Fig. \ref{fig:reconstructedConcFieldLagrangian} shows the numerical reconstruction of the spatial distribution of scalar at t=20, where the time has been normalized by the time scale for advection assumed to be $=1$. In Figure \ref{fig:reconstructedConcFieldLagrangian}(a), corresponding to $Pe=10^{3}$, the effect of stretching-enhanced diffusion can clearly be seen in many parts of the material lamella, especially at the region where the flow field changes its nature from linear to hyperbolic. Higher velocity gradients in that zone cause greater stretching and consequently more diffusion broadening. Different parts of the strip have overlapped to a significant extent over there, destroying concentration gradients, giving an indication that the mixing time had been attained well before. On the other hand, in Figure \ref{fig:reconstructedConcFieldLagrangian}(b), the concentration has not decayed much at parts of the strip away from the center of the vortex whereas segments closer to the vortex eye have just started to overlap, implying that the mixing time has just been reached for $Pe=10^{5}$. 

\subsection{Probability density function of concentration}\label{sec:pdfConc}

An accurate description of the evolution of the probability density function of the scalar is essential in order to understand the mixing dynamics in a Rankine vortex. Numerically, the PDF can be obtained by computing the histogram of the spatial distribution of scalar over the 2-D grid. This is computationally intensive since it requires a full reconstruction on a 2-D grid. However, there is a faster alternative that uses the fact that each strip has a Gaussian transverse variation. This is done by computing the histogram of concentrations for each segment of the strip by integrating normal to the strip in the local co-ordinate system and then adding the contributions from all segments of the strip in order to get the total PDF. Since the PDF is a non-linear function of the concentration, a linear sum computed over the strip will not yield the correct result once coalescence occurs; hence, this method is valid only at initial stages when different parts of the strip do not overlap. In spite of its limited accuracy, this is an important quantity as its deviation from the numerically computed PDF will be a measure of the extent of mixing that has occurred in the flow. 

We attempt at deriving the analytical expression of the PDF of concentration P(c) by making the following change of variables:

\begin{equation}
    P(c;t)dc = C(x;t)dx
\end{equation}

where the concentration PDF P(c;t) is defined so that P(c;t)dc is the probability, or frequency of occurrence, of concentration c at time t in the domain and C(x;t)dx is the mass in an area dx centered at x. Here, we assume a uniform distribution of x over a fixed domain of interest. This allows us to write

\begin{equation}\label{P(c)C(x)}
    P(c) = \frac{C(x)}{\frac{dc}{dx}}
\end{equation}

\begin{figure}
    \centering
    \includegraphics{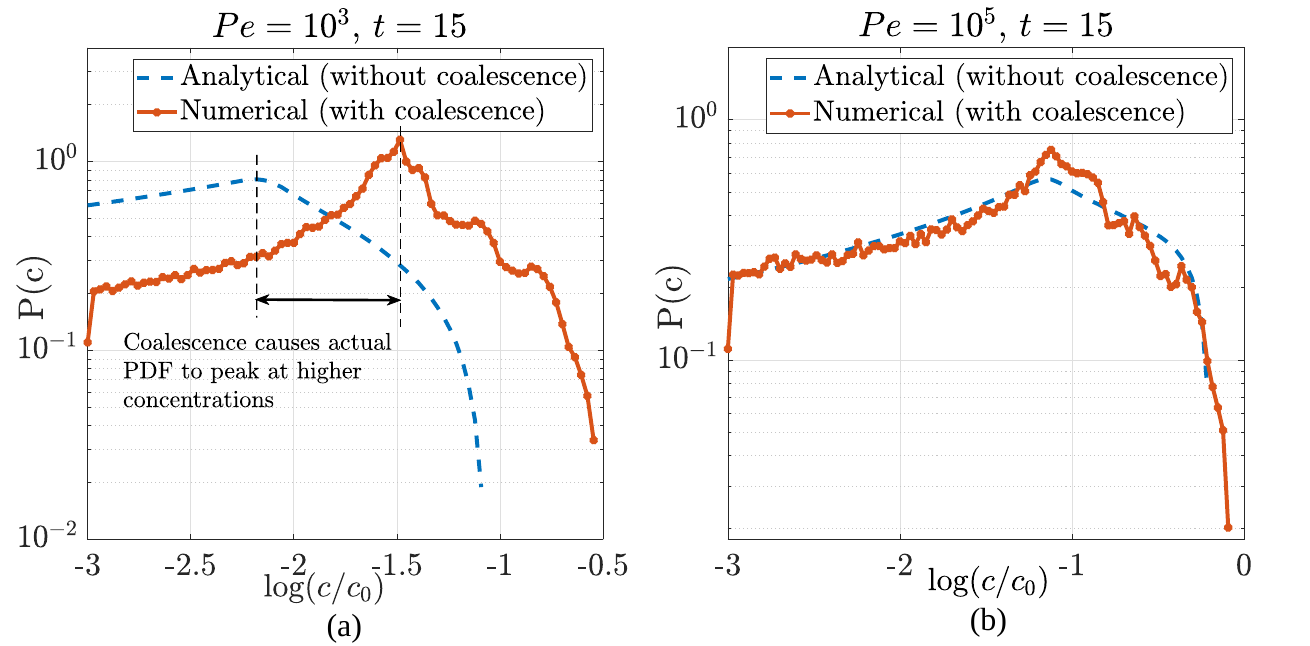}
    \caption{PDF of concentration of scalar for (a) $Pe=10^{3}$ and (b) $Pe=10^{5}$ at t=15.}
    \label{fig:concPdf}
\end{figure}

At an arbitrary time in the simulation, the concentration for a particular segment is given by (\ref{concPerpToStrip}) . Substituting (\ref{concPerpToStrip}) in \ref{P(c)C(x)} above yields

\begin{equation}\label{dc/dx}
    \frac{dc}{dx} = c \frac{-2x}{s_{i}^{2} (1+4\tau_{i})}
\end{equation}

with x $\equiv$ n. By separating the variables in the equation above and integrating from the strip center to a certain distance normal to the strip in the local co-ordinate system, we get the following:

\begin{equation}
    \int_{c_{max}}^{c} \frac{1}{c} dc = \int_{0}^{x} \frac{-2x}{s_{i}^{2} (1+4\tau_{i})} dx 
\end{equation}

where $c_{max} = \frac{1}{\sqrt(1+4\tau_{i}(t))}$. The integration above gives an expression for x which when substituted in (\ref{dc/dx}) gives the final expression for the PDF for the $i^{th}$ segment as

\begin{equation}\label{P(c)}
    P_{i}(c) = A \frac{s_{i} \Delta l \sqrt{1+4\tau_{i}}}{c \sqrt{\ln{(c_{max}/c)}}}
\end{equation}

Note that the normalizing constant A cannot be obtained from this analytical expression since the integral $\int_{-\infty}^{\infty} P(c)dc = 1 $ is divergent at $c=0$ \citep{meunier2010diffusive}. The PDF over the entire domain of interest is calculated by adding the contributions from all such segments (assuming that different parts of the strip do not overlap with each other): $P(c) = B \sum_{i=1}^{n} P_{i}(c)$. We superimpose the PDF obtained from the numerical reconstruction on the PDF derived from analytical considerations and draw a number of interesting inferences regarding the mixing dynamics of the flow from the comparison. 

Figure \ref{fig:concPdf}(a) shows the PDF of the concentration of scalar for $Pe=10^{3}$ at $t=15$. Unlike the PDF for $Pe=10^{5}$, we observe a much faster decay in concentration in this case since diffusion plays a more prominent role in low P\'eclet number flows. The numerically obtained PDF is seen to have a higher frequency of cells in intermediate and high concentration regimes than that in the analytical PDF. This deviation arises due to aggregation of different parts of the tracer which occurs when the strip width, defined by $w_{i}=s_{i}\sqrt{1+4\tau_{i}}$, is of the order of the distance separating adjacent segments of the folded strip. This aggregation or re-connection between different parts of the mixing interface causes the PDF to peak at higher concentrations. Hence, the difference between the actual PDF and the analytical PDF in intermediate and high concentration zones is essentially how the presence of coalescence manifests itself in the global PDF of concentration. On the other hand, the two different versions of the PDF in \ref{fig:concPdf}(b) are a good match with one another because, at that stage of the simulation, the mixing time has not been reached for $Pe=10^{5}$. 

It must be noted that the numerical PDF was obtained by assuming that the concentration at an arbitrary position in space is equal to a simple addition of the concentration contributions from different Gaussian segments. This is a valid construction rule for the global PDF since the Fourier equation which governs the evolution of the tracer is linear, thereby allowing us to superimpose the concentration fields from various sources.

\begin{figure}
    \centering
    \includegraphics[scale=0.96]{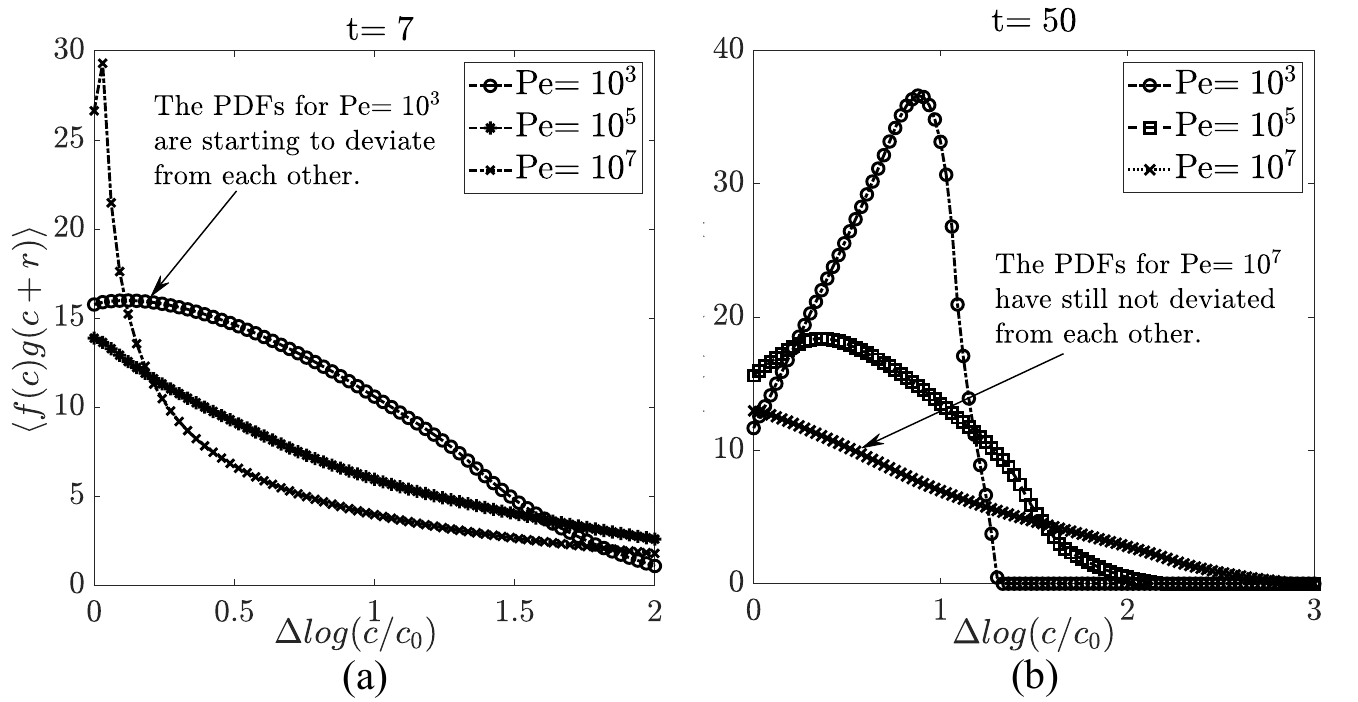}
    \caption{The cross correlation function of the analytical PDF (denoted by $f$) and numerical PDF (denoted by $g$) is shown at t$=7$ and at t$=50$. Even at times as early as t$=7$, the two PDfs start deviating from one another for Pe$=10^{3}$ implying a very fast onset of coalescence relative to the start of the stirring protocol. On the other hand, even at t$=50$, the cross correlation function of the two PDFs peaks around $k=0$ for Pe$=10^{7}$.}
    \label{fig:tCorrelation}
\end{figure}

\begin{figure}
    \centering
    \includegraphics[scale=0.85]{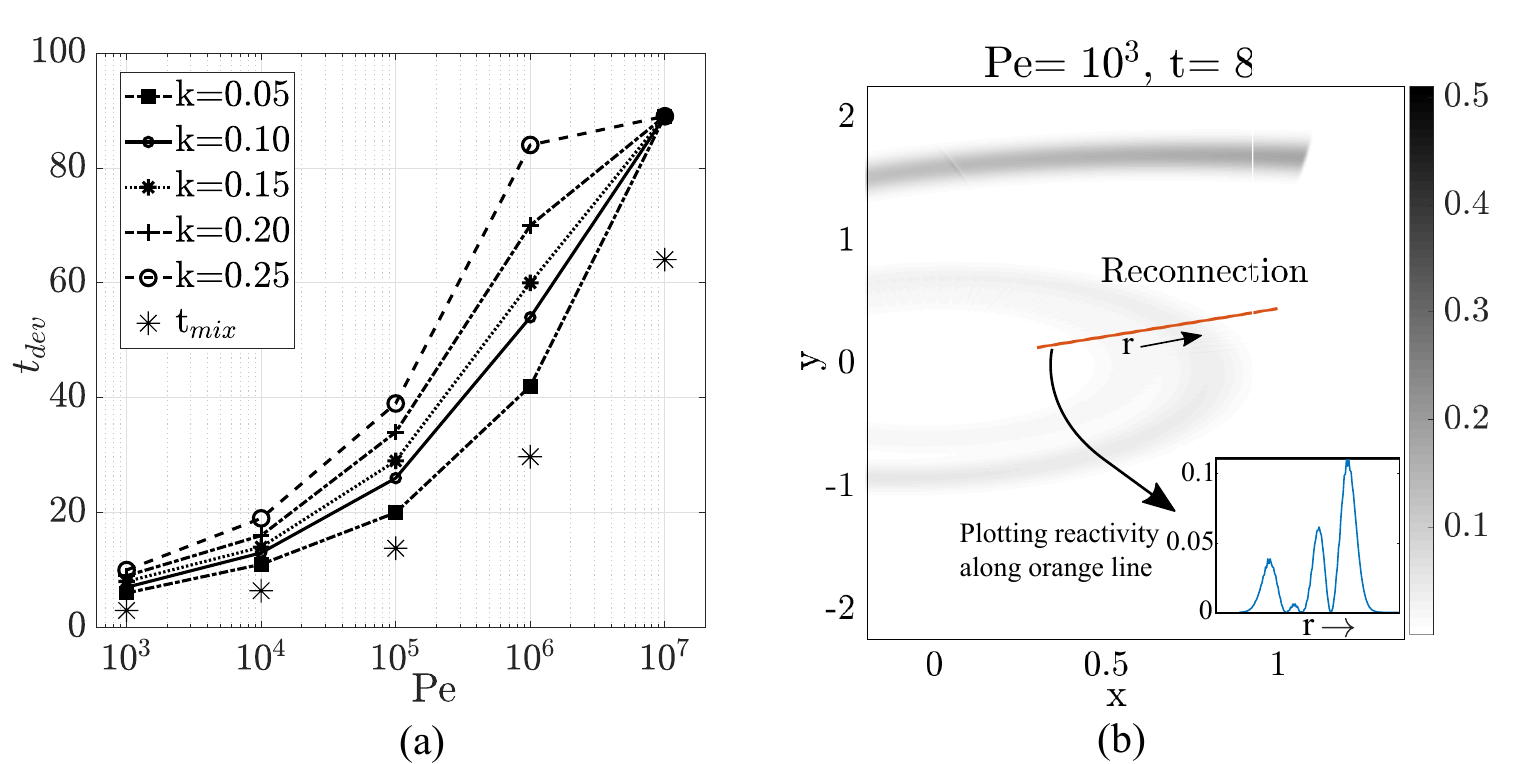}
    \caption{(a) The characteristic time for the analytical and numerical PDFs to deviate from one another is plotted against the P\'eclet number for a range of $k=\Delta$log$(c/c_0)$. (b) A very early onset of coalescence is observed for Pe$=10^{3}$ from the reconnection of different parts of the material interface as highlighted by the orange line. The inset in (b) shows the reactivity along the orange line as a function of the square of the concentration gradient normal to the strip.}
    \label{fig:tDev}
\end{figure}

We carry out a numerical exercise to identify the onset of coalescence across a range of P\'eclet numbers by observing a characteristic time at which the analytical and numerical probability density functions (PDFs) start deviating from each other. In order to draw conclusions about onset of coalescence, we compute the cross correlation function of the two PDFs and observe the peak shifting to higher values of $\Delta$log$(c/c_0)$ with time (See Figure \ref{fig:tCorrelation}). The characteristic deviation time $t_{dev}$ is defined as the time at which the peak of the cross correlation function crosses $\Delta$log$(c/c_0)=k$ where $k$ can be set arbitrarily based on a threshold that the end-user of this algorithm sets for coalescence. Although the definition of $t_{dev}$ involves arbitrarily selecting the critical offset as the measure of coalescence, we see that irrespective of this value, a rough scaling emerges between the deviation time and P\'eclet number. This quantity is shown for a reasonable range of $k$ in Figure \ref{fig:tDev}(a). 

We compare the time at which coalescence sets in with the mixing time and make a few qualitative remarks on the consequences of these events on reactivity. Beyond the mixing time, the maximum concentration at the center of the strip $ c_{max} = c_{0}/\sqrt{1+4\tau}$ decays as $1/t^{3/2}$. The event of mixing time $t_{mix}$ corresponds to the time instant when the strip width, averaged over the entire strip, is at a minimum. Consequently, concentrations gradients and reactivity are at a maximum. However, folding of the strip in the vortex may bring different parts of the material so close to each other that neighbouring concentration zones overlap at approximately $t_{mix}$. Aggregation of different parts of the material lamella smothers concentration gradients and hence effective reactivity is reduced. For very low P\'eclet number flows, such as Pe$=10^{3}$ the deviation time is reached around the same time as the mixing time implying that concentration gradients exist for a very short time interval compared to that in higher P\'eclet number flows. We observe significant coalescence in this case even at $t=7$ as shown in Figure \ref{fig:tDev}(b). Thus, consideration of only the diffusion kinetics of the vortex beyond the mixing time, without any regard for coalescence, will lead to an overestimate of effective reaction rates. We provide a more quantitative treatment of the influence of coalescence on reactivity in the following section. 

\subsection{Impact of coalescence: fast reaction limit}

The radial variation in the azimuthal velocity deforms the material lamella and consequently enhances concentration gradients. This induces a magnified diffusive flux which is larger than that in the absence of any substrate deformation. Diffusion of a decaying scalar causes neighbouring concentration zones to coalesce, thereby, smothering concentration gradients in the flow field, and ultimately leading to homogenization of the solute. The onset of coalescence strongly influences reaction kinetics in mixing-limited flows and this is illustrated using a special case below.  

\begin{figure}
    \centering
    \includegraphics[scale=0.95]{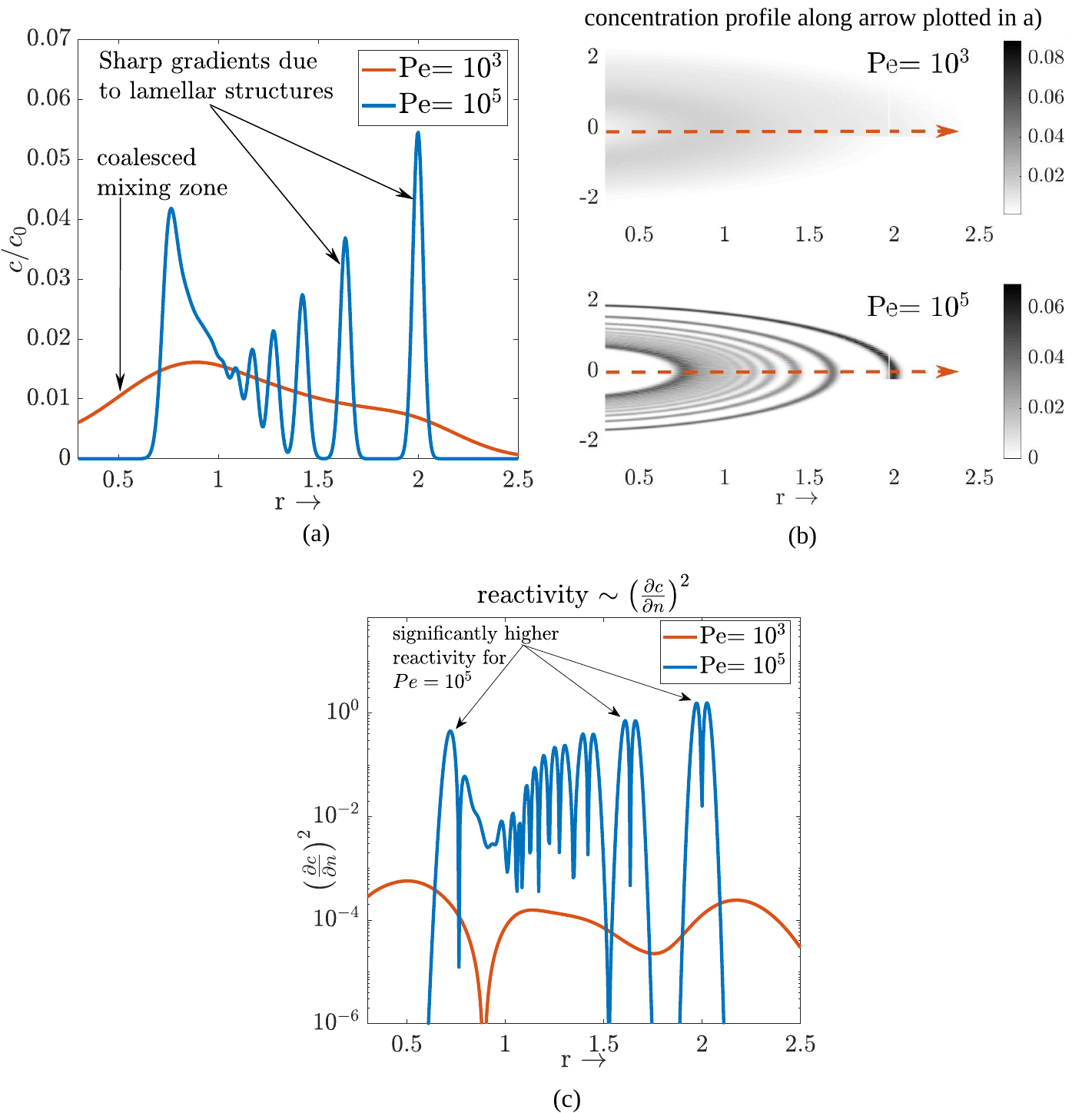}
    \caption{The concentration profile in the radial direction corresponding to two P\'eclet numbers is reconstructed in part b) and is plotted in part a) to reveal the presence of distinct lamellar structures in the flow field. These stretched lamellae provide zones of high concentration gradients and act as fronts of enhanced chemical activity. A plot for the comparative reactivities of two P\'eclet numbers is shown in part c) where we see a reaction rate for $Pe=10^{5}$ which is several orders of magnitude higher than that for $Pe=10^{3}$.}
    \label{fig:consequencesOfCoalescence}
\end{figure}

We consider the case of a bimolecular reaction yielding a single product as follows: $A + B \rightarrow C$. The reactants are denoted as A and B whereas the product is denoted by C and their concentrations are denoted by $a, b$ and $c$ respectively. The equations governing the evolution of the species is given by:

\begin{subequations}\label{eq:reactiveTransport}
  \begin{align}
    \begin{split}
        \frac{\partial a}{\partial t} + \mathbf{v} \boldsymbol{\cdot} \nabla{a} &= D \laplacian{a} - r 
    \end{split} \\[\jot]
    \begin{split}
        \frac{\partial b}{\partial t} + \mathbf{v} \boldsymbol{\cdot} \nabla{b} &=  D \laplacian{b} - r
    \end{split} \\[\jot]    
        \frac{\partial c}{\partial t} &= r
  \end{align}
\end{subequations}

Subtracting (\ref{eq:reactiveTransport} b) from (\ref{eq:reactiveTransport} a) results in 

\begin{equation}
    \frac{\partial q}{\partial t} + \mathbf{v} \boldsymbol{\cdot} \nabla{q} = D \laplacian{q}  
\end{equation}{}

where $q=a-b$ and this resembles the equation governing the transport of a conservative scalar. In situations where the diffusion time scale is much greater than the reaction time scale, A and B are in local equilibrium with the deviation from equilibrium driving product formation \citep{de2005procedure}. We observe that the evolution of reactive species can be modelled by analyzing the transport of a conservative quantity and seek to gain insight into reactivity and product formation from our study of a conservative scalar.  

\begin{figure}
    \centering
    \includegraphics[scale=1.05]{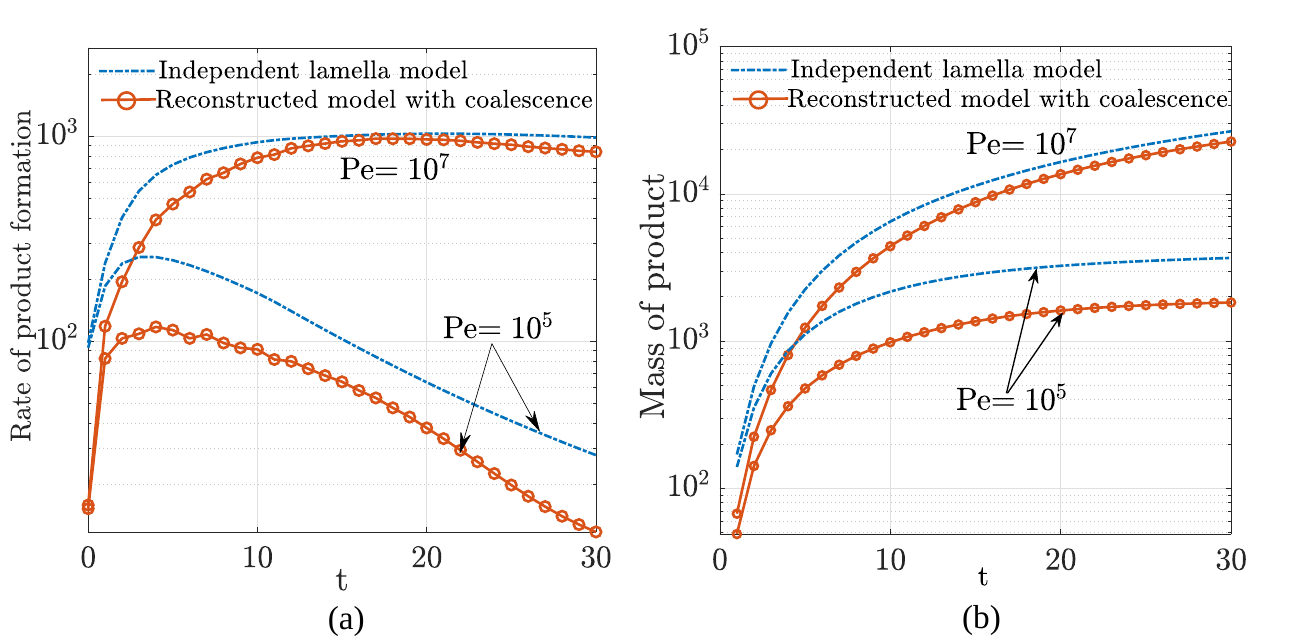}
    \caption{The rate of product formation and the mass of product, both in arbitrary units, is plotted against time for both the independent lamella model and from numerical reconstruction. The hypothesis that the presence of lamellar structures, representing sites of high concentration gradients, enhances chemical reactivity is validated here for fast bimolecular reactions.}
    \label{fig:pdtFormationData}
\end{figure}

It has been shown for mixing-limited bimolecular reactions yielding a precipitate that the rate of reaction is proportional to the square of the spatial gradient in the concentration field $r \sim \left(\frac{\partial c}{\partial n} \right)^{2}$ \citep{de2005procedure}. In low P\'eclet number flows, concentration gradients are smothered faster at regions of diffusive coalescence thereby reducing effective reactivity. These areas of enhanced coalescence and mixing are characterized by an absence of distinguishable lamellae. On the other hand in high P\'eclet number flows, the lamellar structure of the solute is preserved and this creates steeper concentration gradients leading to faster reaction rates. These interfaces with sharp concentration gradients strongly influence subsurface dynamics since they act as hot-spots for biochemical reaction. Such reactive hot-spots create an environment for the development of microorganisms leading to formation of biofilms, govern the reaction dynamics at the interface of saltwater and freshwater bodies and also play a crucial role in chemical injection based remediation of contaminated groundwater. An overlay of the concentration variation and relative reaction rates for two P\'eclet numbers along with the corresponding gray-scale images of the concentration field is shown in Figure \ref{fig:consequencesOfCoalescence} for the Rankine vortex flow.

The time evolution of the mass of the product C is calculated by integrating (\ref{eq:reactiveTransport} c) in the spatial coordinates and in time with an assumption that the reactivity $r = \left(\frac{\partial q}{\partial n} \right)^{2}$ (we set the constant of proportionality to one in order to perform an order of magnitude analysis). Here we observe two trends which are consistent with the hypothesis that the presence of lamellar structures enhances reaction rates. First, higher P\'eclet number flows, having a greater number of distinguishable lamellae than their lower P\'eclet number counterparts, are associated with higher reactivity and consequently greater product formation at a given point in time and second, the predictions, for the same quantities, from the independent lamella model (which does not account for coalescence - see section \ref{sec:pdfConc}) slightly exceed those obtained from a high resolution reconstruction. These results clearly indicate that the discerning feature governing reaction kinetics at an interface is the extent of coalescence or its absence thereof. 

The ubiquity of the Rankine vortex in atmospheric phenomena makes this fundamental study an important contribution to our understanding of mixing dynamics in the atmosphere. Velocity gradients being higher near the eye of the vortex stir the interior of the vortex more vigorously (see Figure \ref{fig:reconstructedConcFieldLagrangian}(a)). Consequently we observe the first occurrences of coalescence over there. These interior regions with greater mixing intensities have a pronounced effect on the transport of reactive species such as ozone, chlorine monoxide and water vapour, consequently having a strong influence on Antarctic ozone hole dynamics. 

\begin{figure}
    \centering
    \includegraphics{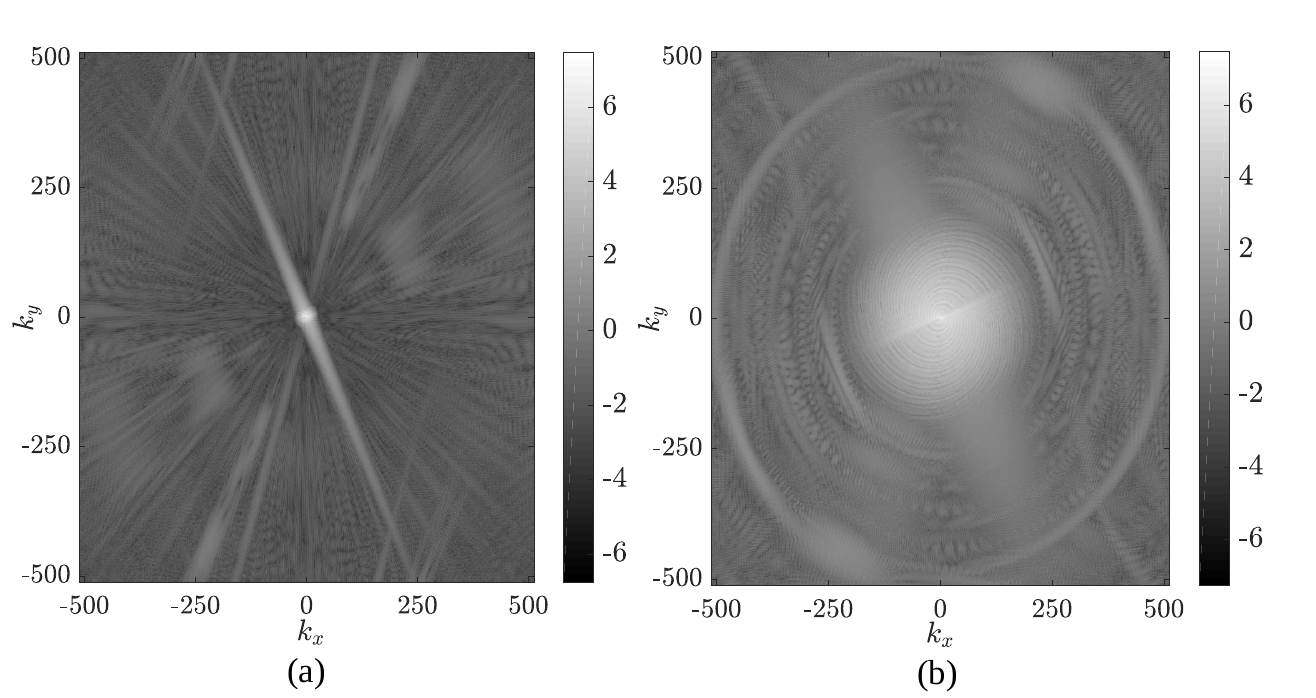}
    \caption{The 2-D spectrum for $Pe=$(a)$10^{3}$ and (b)$10^{5}$, both computed at $t=20$. Higher values of energy can be observed for a wider range of wavenumbers in the case of $Pe=10^{5}$.}
    \label{fig:2DScalarSpectrum}
\end{figure}

Understanding the mixing dynamics in such prototypical laminar flows is the first step towards a mechanistic understanding of transport in reactive flows. In future works, we will quantitatively elucidate the effect of distinguishable lamellae and solute aggregation on reaction kinetics by incorporating a reaction term in the species conservation equation (\ref{eq:advDiffusion}) and a parametric sweep with respect to Damk\"ohler number.

\subsection{Energy spectrum of the scalar}

In this section we discuss the energy spectrum of the scalar advected in the Rankine vortex. The two-dimensional spectrum $\phi(\mathbf{k})$ is equal to the squared modulus of the Fourier transform of the reconstructed scalar field and can be computed numerically with ease by the Fast Fourier Transform (FFT) of the spatial distrbution of the scalar: $\phi(\mathbf{k}) = \abs{c(\mathbf{\tilde{k}})}^{2}$ where $A$ is the area of the flow domain. Figure \ref{fig:2DScalarSpectrum} shows the 2-D spectrum for $Pe=10^{3}$ and $10^{5}$ at $t=20$. The spectrum has a maximum at very low wavenumbers and has a rough circular symmetry to it due to the axisymmetric nature of the flow field. For the same time instant in the simulation, we also observe a greater magnitude in the spectrum for $Pe=10^{5}$ for the same wavenumbers - an indication that the scalar has decayed to a much greater extent in the case of $Pe=10^{3}$. 

The one-dimensional spectrum $\Gamma (k)$ can be obtained by integrating the two-dimensional spectrum $\phi (k)$ on an annulus of width $dk$ and then dividing the result by $dk$ \citep{batchelor1959small}:

\begin{equation}
    \Gamma(k)dk = \int_{k<\mathbf{\abs{k'}}<k+dk}^{} \phi(\mathbf{k'}) d\mathbf{k'}
\end{equation}

\begin{figure}
    \centering
    \includegraphics[scale=1.08]{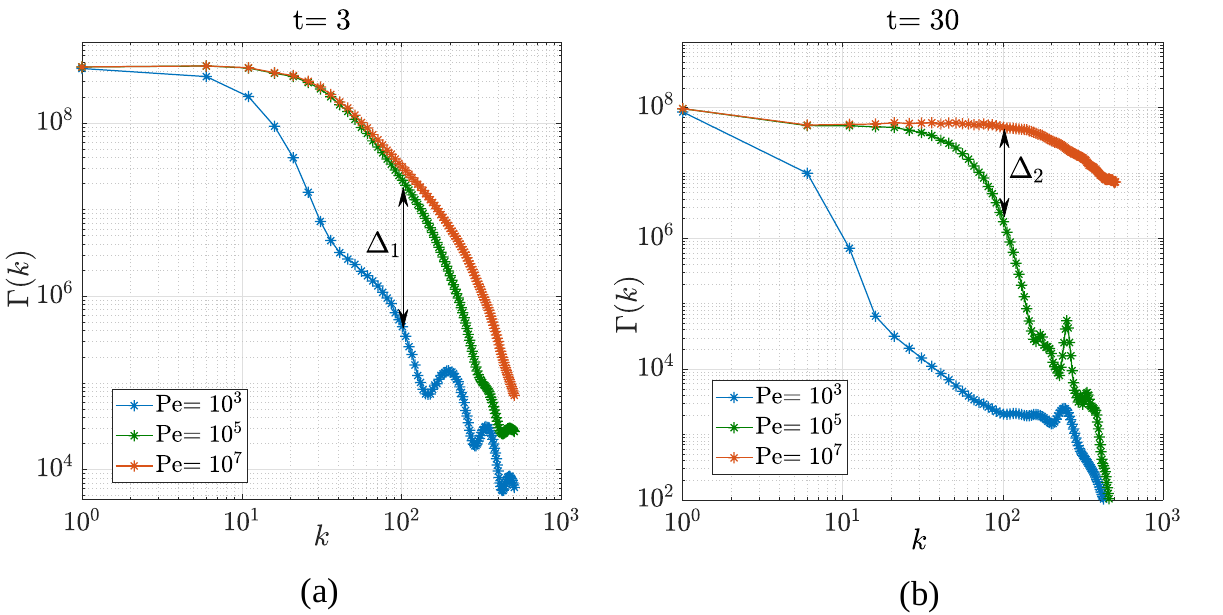}
    \caption{The 1-D energy spectrum is shown for $Pe=10^{3},10^{5},10^{7}$ at $t=3$ and $t=30$ (we are normalizing by the advection time where $t_{advection} = 1$). As the simulation progresses and stretches the strip, we can see the decaying concentration of the scalar manifest itself in the spectrum as a shift towards lower values of energy.}
    \label{fig:1DScalarSpectrum}
\end{figure}

This is shown in Figure \ref{fig:1DScalarSpectrum} for $t=3$ and $t=30$ for $Pe=10^{3}, 10^{5}, 10^{7}$. At $t=3$, the mixing time (defined in ) has been reached for $Pe=10^{3}$. This early onset of coalescence for $Pe=10^{3}$ manifests itself as a significant gap in the spectrum values, denoted by $\Delta_{1}$ in Figure \ref{fig:1DScalarSpectrum}(a)), compared to those of $Pe=10^{5}$ or $10^{7}$. At $t=30$, we observe a difference in the energy spectrum for $Pe=10^{5}$ and $10^{7}$ as well, denoted by $\Delta_{2}$ in Figure \ref{fig:1DScalarSpectrum}(b)) - an indication that the mixing time has been attained for $Pe=10^{5}$. For a decaying scalar, the spectrum shifting to lower magnitudes is typical and this shift is more prominent in the plots for large wavenumbers since we have plotted in logarithmic scale. The numerical computation of the 1-D energy spectrum using an FFT of the reconstructed scalar distribution is more accurate at large wavenumbers since the integration of the 2-D spectrum in an annulus of width $dk$ contains many numerical data at large k. This method yields the spectrum upto $\abs{k}=512$ since the grid for reconstructing the scalar is $1024^{2}$.

\section{Conclusion}

The key signature of coalescence during scalar mixing has been identified as the difference between the actual PDF obtained from numerical simulations and the PDF obtained from theoretical considerations which does not account for the phenomena of strip overlap. Coalescence was observed at very few parts of the tracer for $Pe = 10^5$ even at longer times since the average mixing time for this P\'eclet value had just been reached in the simulation whereas for $Pe = 10^{3}$, coalescence occurs after a very short time interval from the time of injection due to diffusion being more prominent in such low P\'eclet number flows. The construction of the actual global PDF exploits the linear nature of the Fourier equation allowing us to add the contributions of scalar concentration from different segments of the strip of tracer. The comparison of the analytical and numerical PDFs offers insight into the extent of mixing, however, from a numerical viewpoint, efforts must be made to develop techniques which will identify the onset of coalescence without reference to a hypothetical analytical PDF which fails to capture coalescence. 

In the case of low P\'eclet number flows, folding of fluid elements is shown to bring different parts of the material strip closer to each other such that they overlap at the same time as the maximum concentration in the strip starts to decay. This early onset of coalescence leads to an overestimation of reactivity if the dynamics considers only diffusion kinetics without accounting for material aggregation. A formulation is developed by comparing the analytical and numerical PDFs which serves as a guideline for assessing the time of homogenization of solute in such a canonical flow. We do a quantitative analysis of the reaction kinetics in mixing-limited reactive flows and identify the absence of coalescence, or equivalently, the presence of prominent lamellar structures in the flow field as interfaces of enhanced reaction kinetics. This is a problem of great interest to the geophysical community because of the importance of vortex structures and their role in mixing of dispersed tracers. We wish to investigate this problem in greater quantitative detail in the future by including the possibility of reaction in our mathematical formulation of the governing differential equations. 

This work has improved our understanding of the coalescence regime during a mixing protocol of a prototypical flow and essentially arms us with the tools to analyse the mixing behavior of flows in more complicated scenarios. These ideas will also be extended to actual flows occurring in the subsurface where scalar blobs are sheared at different rates due to the heterogeneity of the flow field. This will subsequently build the foundations for analysing flows which are of importance in nature and serve as a thorough precursor to a study of reactive flows.

\appendix
    
\renewcommand{\thefigure}{A\arabic{figure}}
\setcounter{figure}{0}    
    
\section{Lagrangian framework - The diffusive strip method}

In the Lagrangian framework, the modelling technique - Diffusive Strip Method, proposed by Munier and Villermaux \citep{meunier2010diffusive} is integrated with particle tracking simulations to analyse the evolution of a passive scalar. This technique builds on the simplified mathematical framework introduced by Ranz \citep{ranz1979applications}. Passive material strips are advected in the velocity fields given apriori, and their positions are computed kinematically. Diffusion can then be accounted for, independent of advection, by inserting diffusive material segments along the strip. We proceed to discuss the details of the algorithm, starting with the change in the frame of reference necessary in a Lagrangian framework.

\begin{figure}
    \centering
    \includegraphics[scale=1.6]{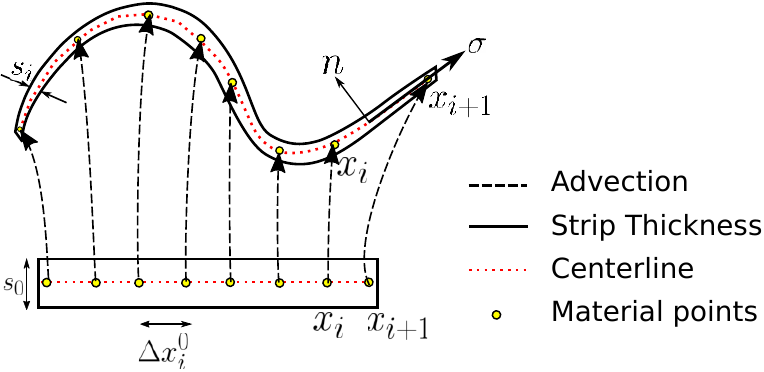}
    \caption{Schematic of tracer representation using a finite number of points at the initial time instant - shown as a straight strip, and at a later time instant - in a deformed configuration.}
    \label{fig:initStripLoc}
\end{figure}

The passive scalar is shown in its initial configuration, as a straight strip of points, in Fig. (\ref{fig:initStripLoc}a) and in a deformed state at a later point in time in Fig.
(\ref{fig:initStripLoc}b). It is convenient to choose a reference frame centered on the strip with $\mathbf{\sigma}$ denoting the strip longitudinal direction locally and \textbf{n} denoting the strip transverse direction locally. 

The tracers are initially separated by a distance  $\Delta x_{i}^{0}$ and have an initial striation thickness of $s_{0}$. As the simulation progresses, the lamella length scale in
the locally transverse direction gets much smaller than that along the strip causing the transverse scalar gradient to be much larger than the scalar gradient along the strip. This allows us to modify the advection-diffusion equation to:

\begin{equation}\label{eq:concGradientPerpToStrip}
    \frac{\partial c}{\partial t} + \frac{n}{s}\frac{ds}{dt}\frac{\partial c}{\partial n} = D\pdv[2]{c}{n}
\end{equation}

We non-dimensionalize this equation by taking the following transformations:
\begin{center}
    $c' = \frac{c}{c_0}$, $t' = \frac{t}{t_{adv}}$, $n'=\frac{n}{L}$ \: and \: $s'=\frac{s}{L}$
\end{center}

Eq. (\ref{eq:concGradientPerpToStrip}) now becomes:

\begin{equation}\label{eq:nonDimConDiff}
    \frac{\partial c'}{\partial t'} + \frac{n'}{s'}\frac{ds'}{dt'}\frac{\partial c'}{\partial n'} = D\frac{t_{adv}}{L^{2}}\frac{\partial^2 c'}{\partial n'^{2}}
\end{equation}

The Ranz transformation \citep{ranz1979applications} converts Eq. (\ref{eq:nonDimConDiff}) to a 1-D diffusion equation (\ref{eq:1D-Diffusion}). Upon taking the Fourier transform of Eq. (\ref{eq:1D-Diffusion}) and using the convolution theorem, we arrive at the following integral equation:

\begin{equation}\label{eq:integralDiff}
    c(\tilde{n},\tau) = \frac{1}{\sqrt{4\pi \tau}} \int_{-\infty}^{\infty} e^{-x^2} e^{-\frac{(\tilde{n}-x)^2}{4\tau}}dx 
\end{equation}

which yields a dimensionless solution for the concentration profile along the local normal orientation. Upon further  simplification the following dimensional solution is obtained:

\begin{equation}\label{concPerpToStrip}
    c(\tilde{n},t) = \frac{c_{0}}{\sqrt{1 + 4\tau_{i}(t)}} exp\left(\frac{-\tilde{n}^{2}}{1 + 4\tau_{i}(t)}\right)
\end{equation}

The sub-routine for reconstructing the spatial distribution of the scalar on a 2-D grid is performed by adding Gaussian ellipses centered at the mid-point of the segment connecting consecutive points on the material strip. 

\begin{equation}\label{eq:reconstructedConc}
    c(\mathbf{x}) = \sum_{i} \frac{c_{0}/1.7726}{\sqrt{1 + 4\tau_{i}(t)}} exp\left(-\frac{[(\mathbf{x-x_{i}})\cdot \mathbf{\sigma_{i}}] ^ {2}}{\Delta l^{2}} -\frac{[(\mathbf{x-x_{i}})\cdot \mathbf{n_{i}}]^{2}}{s_{i}^{2}(1 + 4\tau_{i}(t))} \right)
\end{equation}

Note that it is easier to compute $\mathbf{\sigma_{i}}$ and $\mathbf{n_{i}}$ if we center the ellipses at the mid-point of [$\mathbf{x_{i}},\mathbf{x_{i+1}}$] instead of centering them at the points itself. The factor of 1.7726 is required to account for overestimation of concentration due to overlap of ellipses \citep{meunier2010diffusive}.

\bibliographystyle{jfm}

\end{document}